\documentclass[twocolumn,superscriptaddress,aip,jcp]{revtex4}
\usepackage{amsmath,amssymb, bm}
\usepackage{graphicx,color}
\usepackage{boxedminipage}
\usepackage{epstopdf}

\renewcommand{\d}{\mbox{d}}
\newcommand{\cf}{cf.\,}
\newcommand{\greater}{\mbox{\tiny $>$}}
\newcommand{\lesser}{\mbox{\tiny $<$}}
\newcommand{\lgter}{\mbox{\tiny $\lessgtr$}}

\newcommand{\w}{\omega}
\newcommand{\wti}{\widetilde}
\newcommand{\B}{\mbox{\tiny B}}

\newcommand{\T}{\mbox{\tiny T}}
\newcommand{\tS}{\mbox{\tiny S}}
\newcommand{\SB}{\mbox{\tiny SB}}

\newcommand{\ti}{\tilde}
\newcommand{\nl}{\nonumber \\}
\newcommand{\la}{\langle}
\newcommand{\ra}{\rangle}

\newcommand{\Sec}[1]{Sec.\,\ref{#1}}
\newcommand{\App}[1]{Appendix\,\ref{#1}}
\newcommand{\be}{\begin{equation}}
\newcommand{\ee}{\end{equation}}
\newcommand{\bea}{\begin{eqnarray}}
\newcommand{\eea}{\end{eqnarray}}
\newcommand{\bsube}{\begin{subequations}}
\newcommand{\esube}{\end{subequations}}
\newcommand{\Eq}[1]{Eq.\,(\ref{#1})}
\newcommand{\Eqs}[1]{Eqs.\,(\ref{#1})}
\newcommand{\Fig}[1]{Fig.\,\ref{#1}}

\newcommand{\comments}[1]{}

\begin{document}

\title{Equilibrium and transient thermodynamics:
A unified dissipaton--space approach}

\author{Hong Gong}\thanks{Authors of equal contributions}
\affiliation{Department of Chemical Physics, University of Science and Technology of China, Hefei, Anhui 230026, China}

\author{Yao Wang} \thanks{Authors of equal contributions}
\affiliation{Hefei National Laboratory for Physical Sciences at the Microscale
 and iChEM and Synergetic Innovation Center of Quantum Information
 and Quantum Physics,
 University of Science and Technology of China, Hefei, Anhui 230026, China}

\author{Hou-Dao Zhang} \email{hdz@ustc.edu.cn}
\affiliation{Department of Chemical Physics, University of Science and Technology of China, Hefei, Anhui 230026, China}

\author{Qin Qiao}
\affiliation{Digital Medical Research Center of School of Basic Medical Sciences, Fudan University, Shanghai, 200032, China
}
\affiliation{Shanghai Key Laboratory of Medical Image Computing and Computer Assisted Intervention, Shanghai, 200032, China
}

\author{Rui-Xue Xu}
\author{Xiao Zheng}
\author{YiJing Yan} \email{yanyj@ustc.edu.cn}
\affiliation{Department of Chemical Physics, University of Science and Technology of China, Hefei, Anhui 230026, China}
\affiliation{Hefei National Laboratory for Physical Sciences at the Microscale
 and iChEM and Synergetic Innovation Center of Quantum Information
 and Quantum Physics,
 University of Science and Technology of China, Hefei, Anhui 230026, China}

\date{July 8, 2020 submitted to JCP}

\begin{abstract}
 This work presents a unified dissipaton--equation--of--motion (DEOM)
theory and its evaluations on
the Helmholtz free energy change due to
the isotherm mixing of two isolated subsystems.
One is a local impurity and another is a nonlocal Gaussian bath.
DEOM constitutes a fundamental theory for such open quantum
mixtures.
To complete the theory, we construct also
the imaginary--time DEOM formalism via
an analytical continuation of dissipaton algebra,
which would be limited to equilibrium thermodynamics.
On the other hand, the real--time DEOM deals with
both equilibrium structural and nonequilibrium dynamic properties.
Its combination with the thermodynamic integral formalism
would be a viable and accurate means to both equilibrium
and transient thermodynamics.
As illustrations, we report the numerical results on a spin--boson system,
with elaborations on the underlying anharmonic features, the
thermodynamic entropy versus the
von Neumann entropy, and an indication
of ``solvent--cage'' formation.
Beside the required asymptotic equilibrium properties,
the proposed transient thermodynamics also supports
the basic spontaneity criterion.

\end{abstract}
\maketitle

\section{Introduction}
\label{thintro}

 Entanglement between microscopic systems and
bulk environments is closely related to quantum
dissipative dynamics and thermodynamics properties.
Theoretically this is concerned with
the total system--and--bath composite,
at certain given temperature.
In the thermodynamics nomenclature,
such a total composite constitutes a closed system.
It is in thermal contact with
surrounding reservoir to maintain the constant temperature
scenario.

 Traditionally, quantum dissipation theories
focus only on the reduced system density operator,
$\rho_{\tS}(t)\equiv {\rm tr}_{\B}\rho_{\T}(t)$;
i.e., the bath--subspace trace
over the total composite $\rho_{\T}(t)$.
Analytical solutions
can only be obtained for noninteracting (harmonic) systems,
such as quantum Brownian oscillators.\cite{Hu922843,Xu09074107,%
Cal83374,Gra8487,Gra88115,Hor057325,Hor081161,OCo0615,Cam09392002}
Nonperturbative and numerically exact theories
are also available for anharmonic systems,
with Gaussian--Wick's bath environments.
These include the path integral influence functional
formalism \cite{Fey63118}
and its derivative--equivalence,
the hierarchical--equations--of--motion (HEOM)
approach.\cite{Tan89101,Tan906676,Tan06082001,%
Yan04216,Xu05041103,Xu07031107}
The imaginary--time HEOM formalism
has also been developed for quantum thermodynamics
problems.\cite{Tan14044114,Tan15144110}

 This work is concerned with
the dissipaton--equation--of--motion (DEOM)
theory\cite{Yan14054105,Yan16110306,Zha18780}
and its evaluations on quantum thermodynamics properties.
This theory not only recovers the HEOM formalism,%
\cite{Tan89101,Tan906676,Tan06082001,Yan04216,Xu05041103,Xu07031107}
but also physically identify the original mathematical auxiliary density operators.
The underlying dissipaton algebra
enables DEOM an explicit theory for entanglement
system--and--bath dynamics.\cite{Yan14054105,Yan16110306,Zha18780,Wan20041102}
Demonstrated examples beyond the HEOM evaluations
include the Fano interference,\cite{Zha15024112}
Herzberg--Teller vibronic coupling,\cite{Zha16204109}
quantum transport shot noise spectrums.
\cite{Jin15234108,Jin18043043,Jin20235144}
The recently developed phase--space DEOM theory
enables also the evaluations on various thermal
transport problems, including the dynamical
heat correlation functions.\cite{Wan20041102}

 This paper consists of two major and closely related topics.
The first one is the unified DEOM theory,
with Gaussian--Wick's bath environments in a class of decomposition schemes.
We will present the latter in \Sec{thsec2}
and the resultant DEOM theory in \Sec{thsec3}.
We illustrate the unified theory
with the Fano--spectrum--decomposition (FSD) scheme\cite{Cui19024110,Zha20064107}
that would be a practical choice in
the extremely low--temperature regime.
Note that the unified HEOM formalism with the FSD scheme
had been constructed recently.\cite{Zha20064107}
The unified DEOM theory would facilitate
the evaluations on various system--and--bath entanglement properties
of strongly correlated quantum
impurity systems.\cite{Yan14054105,Yan16110306,Zha18780,%
Zha15024112,Zha16204109,Jin15234108,Jin18043043,Jin20235144,Wan20041102}
To complete this topic, we
present the equilibrium DEOM solutions in \App{thappA},
and further the imaginary--time DEOM in \App{thappB}.
While the imaginary--time formalism
focuses on equilibrium thermodynamics only,
the real--time DEOM accesses also nonequilibrium
and/or transient analogues.
%

 The second topic of this paper is concerned with
quantum evaluations on thermodynamic properties
and the possible extension to transient thermodynamics
problems.
It is worth re-emphasizing that
the total system--and--bath composite
is actually a closed  system in thermodynamics
at constant temperature.
This is exactly the quantum open system
in the literature of quantum dissipation theories.
In \Sec{thsec4}, we present the Second--Law--based
thermodynamic integral formalism on the free--energy change.\cite{Kir35300,Shu71413,Zon08041103,Zon08041104}
This formalism can readily be implemented with the existing
DEOM theory.
Apparently, the equilibrium DEOM solutions presented in \App{thappA}
are the key ingredients for evaluating
those standard thermodynamic variables.
A natural extension would be anticipated to the transient
thermodynamics via
the real--time DEOM evaluations.
We present the numerical results on
a model spin--boson system
and elaborate the underlying physical implications
in \Sec{thsec5}.
Finally, we conclude this paper with \Sec{thconc}.


\section{Decomposition of environments}
\label{thsec2}

\subsection{Prelude}
\label{thsec2A}

 Let us start with total system--plus--bath composite Hamiltonian,
\be\label{HT}
 \hat H_{\T} = \hat H_{\tS}+\hat h_{\B}+\hat H_{\SB}.
\ee
It governs the dynamics of the total composite
$\rho_{\T}(t)$. However, this is
non-isolate but just a closed thermodynamic system
at a given temperature $T$.
The complete characterization requires also
the statistical thermodynamic descriptions
on environments.\cite{Wei12,Yan05187,Zhe121129,Xu18114103,Liu18245}
Throughout this work we set $\hbar=1$ for
the Planck constant and  $\beta\equiv 1/(k_{\B}T)$,
with $k_{\B}$ being the Boltzmann constant.


 Without loss generality, we illustrate the required
descriptions on environments
with the single--dissipative--mode scenario, where
the system--bath coupling reads
\be\label{HSB}
 \hat H_{\tS\B}=\hat Q_{\tS}\hat F_{\B}.
\ee
The hybrid bath mode $\hat F_{\B}$ assumes
linear. This together with noninteracting
$h_{\B}$ constitute a Gaussian environment bath model.
It applies the stochastic force,
$\hat F_{\B}(t)=e^{i\hat h_{\B}t}\hat F_{\B}e^{-i\hat h_{\B}t}$,
on the dissipative system mode $\hat Q_{\tS}$ that can be
is arbitrary.
Note that $\la \hat F_{\B}(t)\ra_{\B}=0$,
in the bare--bath ensemble average,
$\la\hat O\ra_{\B}\equiv{\rm tr}_{\B}
\big(\hat Oe^{-\beta\hat h_{\B}}\big)/{\rm tr}_{\B} e^{-\beta\hat h_{\B}}$.
The interacting bath spectral density acquires
the ensemble averaged expression of\cite{Zhe121129}
\be\label{Jw_def}
 J(\w)\equiv \frac{1}{2}\!\int_{-\infty}^{\infty}\!{\rm d}\w\,
  e^{i\w t}\la[\hat F_{\B}(t),\hat F_{\B}(0)]\ra_{\B}.
\ee
This is an antisymmetric function and often given via
such as the Browian--oscillator or Ohmic models.
Moreover, the influence of a Gaussian environment
at any given temperature is completely characterized by
the interacting bath correlation function:
\be\label{FDT}
 \la\hat F_{\B}(t)\hat F_{\B}(0)\ra_{\B}
=\frac{1}{\pi}\!\int_{-\infty}^{\infty}\!{\rm d}\w\,
\frac{e^{-i\w t}J(\w)}{1-e^{-\beta\w}}.
\ee
This is the bosonic fluctuation--dissipation theorem.\cite{Wei12,Yan05187,Zhe121129}
One can then obtain the Feynman--Vernon
path--integral influence functional formalism.\cite{Fey63118}

 The celebrated HEOM formalism\cite{Tan89101,Tan906676,Tan06082001,%
Yan04216,Xu05041103,Xu07031107}
is the time--derivative equivalent to the path--integral expression.
To achieve a hierarchical coupling structure,
it requires $\la\hat F_{\B}(t)\hat F_{\B}(0)\ra_{\B}$
be decomposed into basis functions, $\{\phi_k(t); k=1,\cdots,K\}$,
that are derivatives--closed,
\be\label{phi_closure}
 \dot\phi_k(t) = -\sum_{j=1}^K \Gamma_{kj}\phi_j(t).
\ee
On the other hand, in line with various optimized HEOM constructions,\cite{Hu10101106,Hu11244106,Din12224103,Din16204110,%
Tan15224112,Dua17214308,%
Din17024104,Ye17074111,Cui19024110,Zha20064107}
one would like to have the expansion expressions of
(denoting $t\geq 0$ hereafter)
\be\label{FF_in_phi}
\begin{split}
 \la\hat F_{\B}(t)\hat F_{\B}(0)\ra_{\B}
 &=\sum_{k=1}^K c_{k}\phi_k(t),
\\
 \la\hat F_{\B}(0)\hat F_{\B}(t)\ra_{\B}
  &=\sum_{k=1}^K c^{\ast}_{\bar k}\phi_k(t)
  =\la\hat F_{\B}(t)\hat F_{\B}(0)\ra^{\ast}_{\B}.
\end{split}
\ee
The second expression is the time--reversal relation.\cite{Wei12,Yan05187,Zhe121129}
The associate index $\bar k$ is defined via
$\phi_{\bar k}(t)\equiv \phi^{\ast}_{k}(t)$ that
is also a basis set function,
$\phi^{\ast}_{k}(t)\in\{\phi_{j}(t);j=1,\cdots,K\}$.
 Convention HEOM constructions exploit
certain sum--over--poles decomposition schemes
on the Fourier integrand of \Eq{FDT}.
All poles engaged are the first--order type.
This results in $\{\phi_k(t)=e^{-\gamma_kt}\}$
as the basis set functions,
with $\Gamma_{kj}=\delta_{kj}\gamma_k$
for the closure relation (\ref{phi_closure}).
While all Matsubara frequencies via Bose function are real,
those exponents via the poles of $J(\w)$ appear
either real or complex conjugate--paired.
This feature holds true for the unified basis set
$\{\phi_k(t)\}$ exploited in \Eq{FF_in_phi},
with the aforementioned associate index $\bar k$.

\subsection{Fano spectrum decomposition approximant}
\label{thsec2B}

 It is well--known that for the Bose function
the Pad\'{e}--spectrum--decomposition (PSD) method
is the best first--order--poles type.\cite{Hu10101106,Hu11244106}
However, in the near--zero--temperature regime, even the PSD
would be too expensive, due to the effective degeneracy there.
The higher--order--poles would be needed.
It is well--known that
a $P^{\rm th}$--order pole is mathematically equivalent to
the degeneracy of $P$ number first--order poles.

 Consider below the Fano--spectrum--decomposition (FSD)
Bose function approximant:\cite{Cui19024110,Zha20064107}
\be\label{FSD0}
\frac{1}{1-e^{-\beta\w}}
\simeq f^{\text{\tiny PSD}}_{N_0}(\w,T_0;T)
 +\sum_{d=1}^D f_{d}(\w, T; T_0, K_d).
\ee
The first term, $f^{\text{\tiny PSD}}_{N_0}(\w,T_0;T)$,
selects $N_0$ number of PSD poles at
a reference temperature, $T_0>T$,
but with the $k_{\B}T/\w$ contribution intact.
The second sum engages individual $K^{\rm th}_d$--order--poles, with
\be\label{FSD1}
 f_d(\w, T; T_0, K_d)
 = \frac{b_d\w/\check\gamma_d}{[(1+\w/\check\gamma_{d})^2]^{K_d}}.
\ee
The dimensionless FSD parameters,
$a_d\equiv 1/(\beta_0\check\gamma_d)$ and $b_d$,
for those practically used values of $T_0/T\gg 1$ and $D$
are well--tabulated.\cite{Cui19024110,Zha20064107}

 The FSD scheme leads to the interacting
bath correlation function the form of
\be\label{FF_FSD0}
\la\hat F_{\B}(t)\hat F_{\B}(0)\ra_{\B}
=\sum_{j=1}^{K_0}\eta_{j}e^{-\gamma_{j}t}+\sum_{d=1}^D C_d(t;K_d).
\ee
The first term, with $K_0=N_0+N_J$, resembles the result of
the convention sum--over--poles scheme
on the Fourier integrand of \Eq{FDT}.
Beside the those via
$f^{\text{\tiny PSD}}_{N_0}(\w,T_0;T)$,
the first terms of \Eq{FSD0},
it engages also $N_J$ number of
first--order poles from the bath spectral density $J(\w)$.
 The second--term in \Eq{FF_FSD0}
arises from the individual $K^{\rm th}_{d}$--order poles
via \Eq{FSD1}, with
\be\label{FF_FSD1}
 C_d(t;K_d) =
 \sum_{j=1}^{K_d}\check\eta_{dj}(\check\gamma_{d}t)^{j-1}
  e^{-\check\gamma_{d} t}
 =C^{\ast}_d(t;K_d)
\ee
The real function nature is highlighted.\cite{Xu07031107,Yan16110306}

 To proceed, we set $K\equiv K_0+K_1+\cdots+K_D$
and recast \Eq{FF_FSD0} with \Eq{FF_FSD1} as \cite{Zha20064107}
\be\label{FF_FSD}
\begin{split}
\la\hat F_{\B}(t)\hat F_{\B}(0)\ra_{\B}
&=\sum_{k=1}^{K} \eta_{k}(\gamma_kt)^{p_k}e^{-\gamma_{k} t},
\\
 \la\hat F_{\B}(0)\hat F_{\B}(t)\ra_{\B}
&=\sum_{k=1}^{K} \eta^{\ast}_{\bar k}(\gamma_kt)^{p_k}e^{-\gamma_{k} t}.
\end{split}
\ee
The first term of \Eq{FF_FSD0} is naturally included in \Eq{FF_FSD},
together with setting
\be\label{pK0}
 p_{k\leq K_0}=0.
\ee
Those $k\in[K_0+1, K]$ in \Eq{FF_FSD}
are further divided into $D$-domains
in relation to the second term of \Eq{FF_FSD0}.
Denote
\be
 \kappa_0\equiv K_0  \ \ \text{and} \ \ \kappa_d \equiv \kappa_{d-1}+K_{d}.
\ee
The index that labels those parameters from \Eq{FF_FSD1} to \Eq{FF_FSD}
is then
\be\label{FSD_index}
 k = k_{dj}\in [\kappa_{d-1}+1,\kappa_{d}];
\ \ d=1,\cdots, D,
\ee
with
\be\label{FSD_para}
 p_{k_{dj}}\equiv j-1,
\ \ \eta_{k_{dj}}\equiv\check\eta_{dj}=\eta^{\ast}_{k_{dj}},
\ \ \gamma_{k_{dj}}\equiv\check\gamma_d=\gamma^{\ast}_{k_{dj}},
\ee
since  $C_d(t;K_d)$ of \Eq{FF_FSD1} is a real function.

\subsection{Dissipatons--decompostion on environment}
\label{thsec2C}

 Equation (\ref{FF_FSD}) has the unified form of \Eq{FF_in_phi},
with the specifications by \Eqs{pK0}--(\ref{FSD_para}).
The basis functions are
\be\label{phi_FSD}
  \phi_k(t)=(\gamma_kt)^{p_k}e^{-\gamma_{k} t};
\ \ k=1,\cdots, K.
\ee
The resultant closure relation, \Eq{phi_closure} is evaluated to be
\be\label{phi_closure_FSD}
 \dot\phi_k(t) = \gamma_{k}[p_k\phi_{k-1}(t)-\phi_k(t)].
\ee

 Turn to the dissipatons--decomposition,
\be\label{F_f}
 \hat F_{\B}=\sum^{K}_{k=1}\hat f_{k}.
\ee
It reproduces \Eq{FF_FSD} with ($t\geq 0$)
\be\label{ff_t}
\begin{split}
 \la\hat f_{k}(t)\hat f_j(0)\ra_{\B}&=\delta_{kj}\eta_k\phi_k(t),
\\
 \la\hat f_j(0)\hat f_{k}(t)\ra_{\B}&=\delta_{kj}\eta^{\ast}_{\bar k}\phi_k(t).
\end{split}
\ee
This highlights the basic features of dissipatons as statistical
quasi--particles, which together with \Eq{FSD_para},
enable \Eq{phi_closure_FSD}
the $\hat f_k$--level correspondence,
\be\label{HeisB}
 (\partial \hat f_k/\partial t)_{\B}
= \gamma_{k}[p_k(\eta_k/\eta_{k-1})\hat f_{k-1}(t)-\hat f_k(t)].
\ee
This holds true when it
is used in evaluating reduced system--subspace properties;
\cf\Eq{diff} and the comments there.
Moreover, the initial values of \Eq{ff_t} are
\be\label{ff0}
\begin{split}
 \la\hat f_{k}\hat f_{j}\ra^{\greater}_{\B}
&\equiv \la\hat f_{k}(0+)\hat f_{j}(0)\ra_{\B}
= \delta_{kj}\delta_{0p_k}\eta_k,
\\
 \la\hat f_{j}\hat f_{k}\ra^{\lesser}_{\B}
&\equiv \la\hat f_{j}(0)\hat f_{k}(0+)\ra_{\B}
= \delta_{kj}\delta_{0p_k}\eta^{\ast}_{\bar k}.
\end{split}
\ee
Equations (\ref{F_f})--(\ref{ff0}) will be exploited
in the construction of the unified DEOM theory
in \Sec{thsec3}.

In contrast to the HEOM formalism that is rooted at
the path integral expression,  DEOM exploits
the Liouville--von Neumann equation,
\be\label{Liou_eq}
 \dot\rho_{\T}(t)=-i[H_{\tS}+ h_{\B}+H_{\tS\B}, \rho_{\T}(t)].
\ee
Define for later use the commutator ($\times$),
forward ($>$) and
backward ($<$) actions of an arbitrary operator $\hat O$ via
\be\label{Algrter}
\begin{split}
 &\quad \hat O^{\times}\rho_{\T}(t)\equiv [\hat O,\rho_{\T}(t)]
\equiv (\hat O^{\greater}-\hat O^{\lesser})\rho_{\T}(t),
\\
 &\hat O^{\greater}\rho_{\T}(t)= \hat O\rho_{\T}(t)
\ \ \text{and} \ \
 \hat O^{\lesser}\rho_{\T}(t)= \rho_{\T}(t)\hat O.
\end{split}
\ee

\section{Unified dissipatons--space theory}
\label{thsec3}

 The dynamics quantities in the DEOM theory are
the dissipaton density operators (DDOs):\cite{Yan14054105,Yan16110306,Zha18780}
\be\label{DDOs}
 \rho^{(n)}_{\bf n}(t)\equiv \rho^{(n)}_{n_1\cdots n_{K}}(t)
\equiv {\rm tr}_{\B}\big[
\big(\hat f_{K}^{n_K}\cdots\hat f_{1}^{n_1}\big)^{\circ}\rho_{\T}(t)\big].
\ee
Here, $n=\sum_{k} n_{k}$, with $n_{k}=0,1,2,\cdots$
being the excitation number of individual dissipaton.
The product of dissipatons inside
the circled parentheses  is irreducible, satisfying
$(\hat f_{k}\hat f_{j})^{\circ}=(\hat f_{j}\hat f_{k})^{\circ}$
for bosonic dissipatons.
Therefore, $\rho^{(n)}_{\bf n}(t)\equiv \rho^{(n)}_{n_1\cdots n_K}(t)$
of \Eq{DDOs},
specifies the configuration of given
$n$--dissipatons excitation.
The reduced system density operator
is just $\rho_{\tS}(t)\equiv\rho^{(0)}_{\bf 0}(t)$.

 By applying \Eq{Liou_eq} with \Eq{HSB} for \Eq{DDOs}, we
immediately obtain [\cf\Eqs{F_f} and (\ref{Algrter})]
\begin{align}\label{DEOM}
 \dot\rho^{(n)}_{\bf n}(t)
&=-i\hat H^{\times}_{\tS}\rho^{(n)}_{\bf n}(t)
  -i\rho^{(n)}_{\bf n}(t;\hat h^{\times}_{\B})
\nl&\quad\,
  -i\big[\hat Q^{\greater}_{\tS}\rho^{(n)}_{\bf n}(t;\hat F^{\greater}_{\B})
  -\hat Q^{\lesser}_{\tS}\rho^{(n)}_{\bf n}(t;\hat F^{\lesser}_{\B})\big].
\end{align}
Here, $H^{\times}_{\tS}\rho^{(n)}_{\bf n}(t)
\equiv [H_{\tS},\rho^{(n)}_{\bf n}(t)]$,
\be\label{DDO_hB}
 \rho^{(n)}_{\bf n}(t;\hat h^{\times}_{\B})
\equiv
  {\rm tr}_{\B} \Big[\big(\hat f_{K}^{n_k}\cdots\hat f_{1}^{n_1}\big)^{\circ}
  \hat h^{\times}_{\B}\rho_{\T}(t)\Big],
\ee
and
\be\label{DDO_FB}
 \rho^{(n)}_{\bf n}(t;\hat F^{\lgter}_{\B}) = \sum_{j=1}^K
 \rho^{(n)}_{\bf n}\big(t;\hat f^{\lgter}_{j}\big).
\ee
The generalized Wick's theorem gives rise
to \cite{Yan14054105,Yan16110306,Zha18780}
\be\label{Wick0}
\rho^{(n)}_{\bf n}(t;\hat f^{\lgter}_{j})
=\rho_{{\bf n}_{j}^+}^{(n+1)}(t)
 +\sum^K_{k=1} n_k \la\hat f_{k}\hat f_{j}\ra^{\lgter}_{\B}
  \rho_{{\bf n}_k^-}^{(n-1)}(t).
\ee
The associated DDO's index ${\bf n}^{\pm}_k$
differs from ${\bf n} \equiv \{n_1,\cdots,n_k\cdots,n_K\}$
at the specified $n_k$ by $\pm 1$.
Together with \Eq{ff0}, we obtain \Eq{DDO_FB}
the expressions,
\be\label{Wick}
\begin{split}
 \rho^{(n)}_{\bf n}(t;\hat F^{\greater}_{\B})
&=\sum^K_{k=1}\left[\rho_{{\bf n}_{k}^+}^{(n+1)}(t)
  +\delta_{0p_k}n_{k}\eta_{k}\rho_{{\bf n}_{k}^-}^{(n-1)}(t)\right],
\\
 \rho^{(n)}_{\bf n}(t;\hat F^{\lesser}_{\B})
&=\sum^K_{k=1}\left[\rho_{{\bf n}_{k}^+}^{(n+1)}(t)
  +\delta_{0p_k}n_{k}\eta^{\ast}_{\bar k}\rho_{{\bf n}_{k}^-}^{(n-1)}(t)\right].
\end{split}
\ee

 The evaluation on $\rho^{(n)}_{\bf n}(t;\hat h^{\times}_{\B})$, \Eq{DDO_hB},
exploits \Eq{HeisB} that turns out to be
the closure relation in the DEOM construction.
Together with
the Heisenberg equation of motion,
 $i(\partial \hat f_k/\partial t)_{\B}=[\hat f_{\B}, \hat h_{\B}]$,
we obtain
\be\label{diff}
 \rho^{(n)}_{\bf n}(t;\hat h^{\times}_{\B})
=i\sum_{k=1}^K n_k\gamma_k
   \bigg[\frac{p_k\eta_k}{\eta_{k-1}}\rho^{(n)}_{{\bf n}^{+,-}_{k-1,k}}(t)
  -\rho^{(n)}_{\bf n}(t)\bigg].
\ee
This is the generalized diffusion equation,\cite{Yan14054105,Yan16110306}
established previously with $p_k=0$,
via the pure exponential--functions decomposition scheme.

 We have completed determined \Eq{DEOM} with
\Eqs{Wick} and (\ref{diff}), with the result
that is identical to
the path--integral--based HEOM formalism.\cite{Zha20064107}
The DEOM theory goes also with
the dissipaton algebra,
in particular the generalized Wick's theorem,
\Eq{Wick}. This enables DEOM a theory for
entangled system--and--bath
dynamics. 
The dissipaton algebra
on the conjugate momentum to $\hat F_{\B}$
is also available.\cite{Wan20041102}
This ingredient can be readily included
in the present unified DEOM theory.

In the absence of external driving fields,
DDOs will eventually become the
thermal equilibrium ones
at the given temperature $T$.
That is $\rho^{(n)}_{\bf n}(t\rightarrow\infty)
=\bar\rho^{(n)}_{\bf n}(T)$,
reads in the generic form of \Eq{DDOs} as
\be\label{rhon_eq}
 \bar\rho^{(n)}_{\bf n}(T) \equiv {\rm tr}_{\B}\big[
 \big(\hat f_{K}^{n_K}\cdots\hat f_{1}^{n_1}\big)^{\circ}
 \rho^{\rm eq}_{\T}(T)\big].
\ee
As elaborated in \Sec{thsec4} or \Eq{Ahyb_integral} with \Eq{HSB_DEOM},
the equilibrium DDOs
dictate the free--energy change in
equilibrium thermodynamics.
To that end, we present in \App{thappA}
an efficient self--consistent--iteration
approach to the steady--state solutions.\cite{Zha17044105}
To complete the theory, we also
construct an imaginary--time DEOM formalism in \App{thappB}.

\section{Equilibrium versus transient thermodynamics}
\label{thsec4}

\subsection{The thermodynamic integral formalism}
\label{thsec4A}

  First of all, the thermodynamic integral formalism
is based on a result of the Second Law that
the Helmholtz free--energy change in an isotherm process
amounts to the \emph{reversible work}.
We focus on the free--energy change of the total system--and--bath
composite before and after hybridization,
\be\label{Ahyb_0}
 A_{\rm hyb}(T) \equiv A(T)- A_0(T).
\ee
We will see that DEOM supports accurate evaluations
on the thermodynamic integrand.
Moreover, it enables the extension of
equilibrium $A_{\rm hyb}(T)$ to the transient
$A_{\rm hyb}(t;T)$ and the related thermodynamic properties,
as elaborated to the end of \Sec{thsec4B};
see also the comments on \Fig{fig2}.

 Let us start with the hybridization
parameter $\lambda$--augmented
total composite Hamiltonian,
\be\label{HT_lambda}
 \hat H_{\T}(\lambda) = \hat H_{\tS}+ \hat h_{\B}+\lambda \hat H_{\tS\B}.
\ee
A reversible process is now mathematically
described  with the smooth varying the hybridization parameter
from $\lambda=0$ to $\lambda=1$.
The \emph{differential reversible work},
$\delta w_{\rm rev}(\lambda)$
performed in region of $[\lambda,\lambda+\d\lambda]$,
assumes the following expressions that will
be elaborated soon later.
\be\label{delWrev}
 \delta w_{\rm rev}(\lambda)
 =\lambda^{-1}\la\hat H_{\tS\B}\ra_{\lambda}\d\lambda
 = {\rm Tr}[\hat H_{\tS\B}\rho^{\rm eq}_{\T}(T;\lambda)]\d\lambda,
\ee
where
\be\label{HSB_lambda_ave}
 \la \hat H_{\tS\B}\ra_{\lambda} \equiv
 {\rm Tr}[(\lambda\hat H_{\tS\B})\rho^{\rm eq}_{\T}(T;\lambda)].
\ee
The hybridization free--energy is then
\be\label{Ahyb_integral}
  A_{\rm hyb}(T) = \int^{1}_0\! \delta w_{\rm rev}(\lambda)
 = \int^{1}_0\! \frac{\d\lambda}{\lambda} \la\hat H_{\tS\B}\ra_{\lambda}.
\ee
This is the thermodynamic integral formalism. \cite{Kir35300,Shu71413,Zon08041103,Zon08041104}
Its classical implementation is available
in many molecular dynamics packages and used
for such as the solvation
free--energy predictions.\cite{Phi051781,Sal13198,Abr1519}

 Equation (\ref{HSB_lambda_ave}) highlights the fact
that $\la\hat H_{\tS\B}\ra_{\lambda}$ is just the $\lambda$--augmented
equivalence to the original $\la\hat H_{\tS\B}\ra$ where $\lambda =1$.
This identification
enables all methods on $\la\hat H_{\tS\B}\ra$,
either theoretical or numerical,
are readily applicable for
quantum thermodynamics evaluations.
In particular, for $\hat H_{\SB}=\hat Q_{\tS}\hat F_{\B}$,
by using \Eq{Wick}, we obtain
\be\label{HSB_DEOM}
 \la\hat H_{\tS\B}\ra_{\lambda}
=\lambda\sum_{k}{\rm tr}_{\tS}\big[\hat Q_{\tS}
  \bar\rho^{(1)}_{k}(T;\lambda)].
\ee
Here, $\bar\rho^{(1)}_{k}(T;\lambda)\equiv
\bar\rho^{(1)}_{{\bf 0}^{+}_k}(T;\lambda)$,
a $\lambda$--augmented DDO at thermal equilibrium,
with the generic \Eq{rhon_eq} but
\be\label{rhon_eq_lambda}
 \bar\rho^{(n)}_{\bf n}(T;\lambda)
\equiv {\rm tr}_{\B}\big[
 \big(\hat f_{K}^{n_K}\cdots\hat f_{1}^{n_1}\big)^{\circ}
 \rho^{\rm eq}_{\T}(T;\lambda)\big].
\ee

\subsection{General remarks}
\label{thsec4B}

 It is worth elaborating the above formalism further.
Let us verify first that
the hybridization free--energy, $A_{\rm hyb}(T)$,
is independent of the sign of $\lambda$.
 Note that the total composite density operator
depends on the system--bath coupling strength
with all--even orders.
In other words, $\rho_{\T}(t;-\lambda)=\rho_{\T}(t;\lambda)$,
including the thermal equilibrium values,
$\rho^{\rm eq}_{\T}(T;-\lambda)=\rho_{\T}(t\rightarrow\infty;\lambda)$.
Consequently, $\la\hat H_{\tS\B}\ra_{-\lambda}=-\la\hat H_{\tS\B}\ra_{\lambda}$
via \Eq{HSB_lambda_ave} is an odd function of $\lambda$.
The resultant $A_{\rm hyb}(T)$ via integral \Eq{Ahyb_integral},
with the integrand $\lambda^{-1}\la\hat H_{\tS\B}\ra_{\lambda}$,
does not depend on the sign of $\lambda$.

 Next, we revisit the reversible work, \Eq{delWrev},
with respect to the First Law about
internal energy with
the form of $U = {\rm Tr}(\hat H_{\T}\rho^{\rm eq}_{\T})$
and the resultant
\be\label{eq11}
 \d U = \delta w + \delta q
= {\rm Tr}[(\d\hat H_{\T})\rho^{\rm eq}_{\T}]
 +{\rm Tr}[\hat H_{\T}(\d\rho^{\rm eq}_{\T})].
\ee
It is noticed that the work $\delta w$ and heat
$\delta q$ arise from the two distinct variations,
$\d\hat H_{\T}$ and $\d\rho^{\rm eq}_{\T}$,
respectively.
For the reversible hybridizing process
described with \Eq{HT_lambda},
we have $\d\hat H_{\T}(\lambda) =\hat H_{\tS\B}\d\lambda$ and
the resultant
$\delta w_{\rm rev}(\lambda)
 ={\rm Tr}\big\{[\d\hat H_{\T}(\lambda)]\rho^{\rm eq}_{\T}(T;\lambda)\big\}
$
the expression (\ref{delWrev}).
On the other hand, the last term in \Eq{eq11}
describes the heat, as it arises
from the variation of distribution.

 Finally, in contact with the Third Law,
one would have
\be\label{3rd_law}
 \lim_{T\rightarrow 0}\frac{\partial\rho^{\rm eq}_{\T}(T)}
 {\partial T}=0.
\ee
This is true regardless the details of the system-bath coupling,
as long as there are no structural and geometric inhomogeneities.
In this case, there would be no degeneracy
associated with $\rho^{\rm eq}_{\T}=e^{-\beta\hat H_{\T}}/Z_{\T}$
when $T \rightarrow 0$.
This together with \Eq{Ahyb_integral}--(\ref{rhon_eq_lambda})
would suggest further the hybridization entropy
the limit of
\be\label{S_T=0}
 \lim\limits_{T \to 0} S_{\rm hyb}=
 -\lim\limits_{T \to 0} \frac{\partial A_{\rm hyb}}{\partial T} =0.
\ee

 Apparently, by using
\Eqs{Ahyb_integral}--(\ref{rhon_eq_lambda}),
one can perform accurate DEOM/HEOM evaluations on
thermodynamic properties.
 We evaluate $\bar{\bm\rho}(T;\lambda)
\equiv \big\{\rho_{\bf {n}}^{(n)}(T;\lambda)\big\}$
progressively, with noting that
$\bar\rho^{(0)}_{\bf 0}(T;0)=e^{-\beta H_{\tS}}/Z^{\tS}_0$
and $\bar\rho_{\bf {n}}^{(n>0)}(T;0)=0$.
Then using the known $\bar{\bm\rho}(T;\lambda)$
as the initial values for calculating
$\bar{\bm\rho}(T;\lambda+\delta\lambda)$
via either the real--time ($t\rightarrow\infty$) propagation (\Sec{thsec3})
or steady--state--solutions (\App{thappA}) approach.
Other accurate methods are primarily
concerned with zero--temperature and static case.
These include the quantum Monte Carlo approach,
\cite{Hir862521, Gul11349}
density matrix renormalization group
\cite{Whi922863,Vid03147902}
and numerical renormalization group
\cite{Wil75773,Bul08395} methods.

 On the other hand, DEOM is a dynamics theory and
naturally applicable to transient thermodynamics with the thermodynamics
integral, \Eq{Ahyb_integral}.
In particular, we replace the equilibrium $\bar\rho^{(1)}_{k}(T;\lambda)$
in the thermodynamics
integrand, \Eq{HSB_DEOM}, with the time--dependent $\rho^{(1)}_{k}(t;T;\lambda)$.
Its evaluation involves the time propagation of the $\lambda$--augmented
counterpart to \Eq{DEOM}. The resultant \Eq{Ahyb_integral} gives
rise to $A_{\rm hyb}(t;T)$ that asymptotically approaches to the equilibrium hybridization free--energy when $t\rightarrow\infty$.

\section{Numerical demonstrations}
\label{thsec5}

For numerical demonstrations, we consider a spin-boson model,
in which system Hamiltonian and dissipative mode are [\cf\Eq{HSB}]
\be\label{H_S}
 \hat H_{\tS}={\varepsilon}\hat{\sigma}_z+\Delta\hat\sigma_x
\ \ \,\text{and}\ \ \,
\hat{Q}_{\tS}=\hat{\sigma}_{z},
\ee
respectively. Here, $\{\hat{\sigma}_{i}\}$ are the Pauli matrices,
$\varepsilon$ is the energy bias parameter and $\Delta$ the interstate coupling.
 Adopt for the bath spectral density the Drude model,
\be\label{Jw}
J(\w)=\frac{\eta\gamma\w}{\w^2+\gamma^2},
\ee
where $\eta$ and $\gamma$ are the system-bath coupling strength and
bath cut--off frequency, respectively.
In all the simulations below, we set $\varepsilon=0.5\Delta$,
$\gamma=4\Delta$ and $\eta=0.5\Delta$.


\begin{figure}
\includegraphics[width=0.4\textwidth]{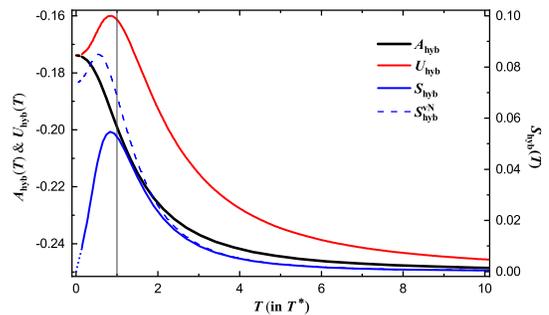}
\caption{
 The hybridization free--energy $A_{\rm hyb}$ (in
  unit of $\Delta$) and the hybridization
  entropy $S_{\rm hyb}$ (in unit of $k_{\B}$).
We set $\varepsilon=0.5\Delta$,
$\gamma=4\Delta$ and $\eta=0.5\Delta$.
Temperature $T$ is in unit of
$T^{\ast}$, \Eq{Tstar}.
See text for the details.
}\label{fig1}
\end{figure}

  Figure \ref{fig1} depicts the resultant hybridization
free--energy $A_{\rm hyb}(T)$ (black),
internal energy $U_{\rm hyb}(T)$ (blue)
and entropy $S_{\rm hyb}(T)$ (red).
The observations and discussions are as follows:
(\emph{i})  $A_{\rm hyb}(T) \leq 0$
that indicates the hybridization is spontaneous.
It decreases monotonically and is flat off
to the value of  $A_{\rm hyb}(T\rightarrow \infty)\approx -0.247\Delta$;
(\emph{ii})
The spontaneous process in study is both
energetically and entropically favored,
as the observed $U_{\rm hyb}(T)<0$
and $S_{\rm hyb}(T)>0$;
(\emph{iii})
There is a characteristic temperature, $T_\text{max}$,
at which $U_{\rm hyb}(T)$ reaches its maximum;
(\emph{iv}) It is noticed that for quantum Brownian oscillators
that are purely harmonic, neither $U_{\rm hyb}$
nor $S_{\rm hyb}$ exhibits the turnover behaviors.\cite{OCo0615,Hor057325,Hor081161}
One can therefore conclude that
the observed turnovers in \Fig{fig1}
are an anharmonic characteristic;
(\emph{v})
However, the quantification
on $T_\text{max}$ is generally complicated.
It involves the interplay between
the solvent induced reorganization energy,
the fluctuation time, and how they effectively
participate in the thermodynamic integration.
Nevertheless, as inferred from \Eq{H_S} with \Eq{Jw},
it would be rather reasonable to have
$T_\text{max}\approx \{[(\varepsilon-\eta/2)^2
+\Delta^2]/\pi\}^{\frac{1}{2}}$;
(\emph{vi}) Adopt for the unit of temperature in \Fig{fig1} rather
a bare--system property,
\be\label{Tstar}
 T^{\ast} = \frac{2}{\tau^{\ast}},
\ \ \, \text{with}\ \ \,
\tau^{\ast}\equiv \frac{2\pi}{\Omega_{\tS}}
= \frac{2\pi}{2\sqrt{\varepsilon^2+\Delta^2}}.
\ee
Here, $\Omega_{\tS} \equiv 2\sqrt{\varepsilon^2+\Delta^2}$
is the characteristic frequency
of the bare two--level $\hat H_{\tS}$ of \Eq{H_S};
thus, $\tau^{\ast}\equiv 2\pi/\Omega_{\tS}$ denotes the periodic time.
The corresponding imaginary time analog is then
$1/T^{\ast} = \tau^{\ast}/2$.
In fact, $T^{\ast}$ does
nicely locate the maximum on the
local--energy difference,
$E^{\rm hyb}_{\tS}(T)\equiv {\rm tr}_{\tS}[\hat H_{\tS}\rho^{\rm eq}_{\tS}(T)]
  -{\rm tr}_{\tS}[\hat H_{\tS}\rho^{\rm eq}_{\tS;\lambda=0}(T)]$,
where $\rho^{\rm eq}_{\tS;\lambda=0}(T)$ is
the unmixed counterpart to $\rho^{\rm eq}_{\tS}(T)$;
(\emph{vii}) In general, $E^{\rm hyd}_{\tS}(T)\neq U_{\rm hyb}(T)$
and their maximum temperatures, $T^{\ast}$ and $T_{\rm max}$,
are also different.
The thermalization underlying the internal energy $U_{\rm hyb}(T)$
is rather complicated, as mentioned early.
The sign of $T_{\rm max}-T^{\ast}$
can be either positive or negative
[not shown here but would have
been implied in (\emph{v}) above].

 Included in \Fig{fig1} is also the
von Neumann entropy counterpart (dash--blue),
\be\label{Shyb_vN}
 S^{\rm vN}_{\rm hyb}(T) \equiv S_{\text {vN}}[\rho^{\rm eq}_{\tS}(T)]
  - S_{\text {vN}}[\rho^{\rm eq}_{\tS;0}(T)],
\ee
with $S_{\text {vN}}[\rho]\equiv -k_{B}{\rm tr}_{\tS}(\rho\ln\rho)$.
This is evaluated at the reduced system density operator in the mixture
at equilibrium, $\rho^{\rm eq}_{\tS}(T)$,
versus its unmixed counterpart, $\rho^{\rm eq}_{\tS;\lambda=0}(T)$.
It is observed that
\be\label{Scmp}
 S_{\rm hyb}(T) \leq S^{\text {vN}}_{\rm hyb}(T).
\ee
While the equal sign holds only for either isolated systems or the $T\rightarrow \infty$ limit, the inequality is general, regardless
whether there is anharmonicity or not.\cite{OCo0615,Hor057325,Hor081161}
The physical picture is as follows.
In contrast to the thermodynamic entropy,
the von Neumann entropy engages the reduced description,
using only $\rho^{\rm eq}_{\tS}(T)={\rm tr}_{\B}\rho^{\rm eq}_{\T}(T)$.
This local system property alone
contains little information on
the system--and--bath entanglement.
The loss of mutual information is responsible
for the fact that
$S_{\text {vN}}(T)$ is always greater than the thermodynamic entropy.
The reduced system density operator $\rho^{\rm eq}_{\tS}(T)$
intrinsically describes a mixed--state property even at $T=0$.
Consequently, $S_{\text {vN}}(T= 0)>0$,
due to the aforementioned information loss.
On the other hand, $S_{\rm hyb}(T\rightarrow 0)=0$,
as anticipated via the Third Law of Thermodynamics.
Nevertheless, the DEOM--evaluation
remains challenge in the zero--temperature limit
and the dot--part of the $S_{\rm hyb}(T)$
curve in \Fig{fig1} is an extrapolation.

 By far we have exploited the equilibrium
DDOs, $\{\bar\rho^{(n)}_{\bf n}(T;\lambda)\}$ of \Eq{rhon_eq_lambda},
and evaluated the hybridization free--energy, $A_{\rm hyb}(T)$,
\Eq{Ahyb_integral} with \Eq{HSB_DEOM}.
It is noticed that the imaginary--time HEOM formalism
for ${\rm tr}_{\B}e^{-\tau H_{\T}}/{\rm tr}_{\B}e^{-\beta h_{\B}}$
had been constructed by Tanimura.\cite{Tan14044114,Tan15144110}
Reported there were also for a spin--boson complex
the numerically accurate $U_\text{hyb}(T)+U^{\tS}_0(T)$
and $S_\text{hyb}(T)+S^{\tS}_0(T)$.
Here, $U^{\tS}_0(T)={\rm tr}_{\tS}[\hat H_{\tS}\rho^{\rm eq}_{\tS;\lambda=0}(T)]$
and $S^{\tS}_0(T)=S_{\text {vN}}[\rho^{\rm eq}_{\tS;0}(T)]$
are the unmixed bare--system contributions.
None of the above two sum--quantities shows
turnover behavior.
Nevertheless, the above type of quantities
does not describe the thermodynamic isotherm process
in a closed system,
whereas the hybridization part does.
It would be prefer to have the imaginary--time DEOM
for the hybridization partition function; see \App{thappB}.

\begin{figure}
\includegraphics[width=0.4\textwidth]{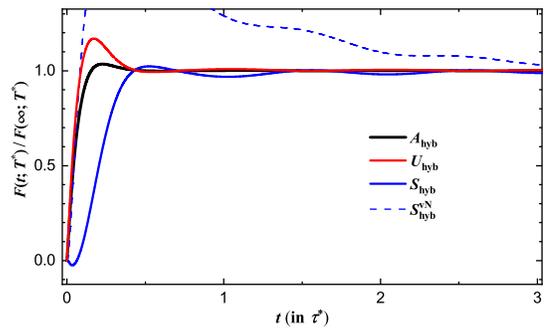}
\caption{The transient thermodynamics of the spin-boson model
of \Fig{fig1}, at $T=T^{\ast}$.
Each of them is scaled by its own $F(t\rightarrow\infty)=F^{\rm eq}$,
with $(A^{\rm eq}_{\rm hyb},U^{\rm eq}_{\rm hyb})/\Delta=(-0.199,-0.161)$
and $(S^{\rm eq}_{\rm hyb},S^{\rm vN; eq}_{\rm hyb})/k_{\B}
=(0.053,0.070)$,
respectively.
}\label{fig2}
\end{figure}


  As mentioned to the end of \Sec{thsec4B},
we can extend the real--time DEOM--evaluation
to the transient $A_{\rm hyb}(t;T)$.
Moreover, let us treat $A\equiv U-TS$
as the mathematical definition.
Together with $S=-(\partial A/\partial T)$,
we will recover all involving equilibrium thermodynamic values
when $t\rightarrow\infty$.
Figure \ref{fig2} reports the resultant
time--dependent functions, $F(t)/F(\infty)$,
for the same four physical quantities in \Fig{fig1}
at $T=T^{\ast}$.
The total composite system was
initially $(t<0)$ in the unhybridized equilibrium state,
prior to the switch on of $H_{\SB}$.
The reported time $t$ is in unit of
$\tau^{\ast}$ [\Eq{Tstar}] that
denotes the oscillatory period of the bare system.
The observations and discussions are as follows:
(\emph{i}) $A_{\rm hyb}(t;T) \leq 0$
would imply its applicability to the
characterization of spontaneous
processes that in principle evolve in time;
(\emph{ii}) Physically,
the internal energy is a time--dependent variable.
It together with (\emph{i})
would further imply that $A_{\rm hyb}\equiv U_{\rm hyb}-TS_{\rm hyb}$
and $S_{\rm hyb}=-(\partial A_{\rm hyb}/\partial T)$
remain valid for transient processes;
(\emph{iii})
It is known that in a thermodynamics
closed system, the entropy change,
$S_{\rm hyb}=-(\partial A_{\rm hyb}/\partial T)$,
can be either positive or negative.
The observed $S_{\rm hyb}(t)<0$ in the short--time region
may indicate the formation of a ``solvent--cage''
surrounding the local two--level impurity in study.
The normal thermalization becomes dominant afterward,
resulting in $S_{\rm hyb}(t>t_\text{cage})>0$.
The observed oscillatory behavior is purely
dynamics in nature;
(\emph{iv}) In contrast, the von-Neumann entropy $S^{\text {vN}}_{\rm hyb}$
is an information measure, which is always
nonnegative.
The inequality $S_{\rm hyb}(t;T) < S^{\text {vN}}_{\rm hyb}(t;T)$
remains true in the time--dependent scenario,
following the same discussions on \Eq{Scmp}.


\section{Concluding remarks}
\label{thconc}

  In summary, we have presented a comprehensive account
on a unified DEOM theory, followed by its accurate evaluations
on both the equilibrium and transient thermodynamics functions.
Our results are applicable to
for arbitrary systems with Gaussian bath environments.
Considered in the unified DEOM theory is a class of polynomial--exponential
bath dissipaton decomposition scheme, \Eq{FF_FSD} or
\Eq{phi_FSD} with \Eq{phi_closure_FSD}.
This includes the recently proposed FSD for an efficient
sum--over--poles expansion
of Bose/Fermi function in the extremely
low--temperature regime.\cite{Cui19024110,Zha20064107}
It is worth re-emphasizing that the DEOM theory consists of not only
\Eq{DEOM} with \Eqs{Wick} and (\ref{diff}),
but also the phase--space dissipaton algebra.\cite{Yan14054105,Yan16110306,Zha18780,Wan20041102}
The resultant DEOM--space quantum mechanics
does unravel the entangled system--and--bath dynamics.\cite{Yan14054105,Yan16110306,Zha18780,Wan20041102}
 Moreover, as shown in \App{thappB},
one can exploit the analytical continuation
dissipaton technique to
the imaginary--time DEOM.
This formalism is numerically preferred
for equilibrium thermodynamics. 

 Alternative approach is the system--bath coupling strength
$\lambda$--integral formalism, \Eq{Ahyb_integral}.
The involving integrand, \Eq{HSB_DEOM},
can be accurately evaluated with the real--time DEOM theory.
This enables the study of both equilibrium and transient thermodynamics.
It is shown that the latter would preserve 
the standard thermodynamic relations
including the spontaneity criterion.
Numerically accurate results are exemplified
with a model spin--boson mixture in \Sec{thsec5}.
This simple system carries two basic characteristics.
One is the anharmonicity and another is
the formation of bosonic solvent--cage surrounding
the local spin impurity.
Interestingly, these two characteristics
are signified in the specified equilibrium
and transient thermodynamics functions,
as reported in \Fig{fig1} and \Fig{fig2},
respectively.

\section*{Data Availability}
The data that support the findings of this study are available from the corresponding author upon reasonable request.

\begin{acknowledgments}
The support from
the Ministry of Science and Technology
(Nos.\ 2016YFA0400900, 2016YFA0200600 and 2017YFA0204904) 
the Natural Science Foundation of China
(Nos.\ 21633006, 21703225 and 21973086)
is gratefully acknowledged.
%

\end{acknowledgments}


\begin{appendix}

\section{Steady-state solutions to DEOM}
\label{thappA}

 This appendix is concerned with the
equilibrium DDOs, $\{\bar\rho^{(n)}_{\bf n}(T)\}$ of \Eq{rhon_eq}.
These are the steady--state solutions,
to the hierarchical dynamics, \Eq{DEOM}, with \Eqs{Wick} and (\ref{diff}).
The standard choices for solving high-dimension linear equations
are the Krylov subspace methods.\cite{Saa03}
However, due to the hierarchy structure in study,
the preferred choice would be
the self--consistent iteration (SCI) method
developed by Zhang \emph{et al}.\cite{Zha17044105}

For bookkeeping we denote
\be\label{calLn_def}
\begin{split}
  {\cal L}^{(n)}_{\bf n}
&\equiv iH^{\times}_{\tS} + \sum_{k} n_k\gamma_k,
\ \ \
  \Lambda_k \equiv \gamma_k\frac{p_k\eta_k}{\eta_{k-1}} ,
\\
 {\cal A} &\equiv \hat Q^{\times}_{\tS}
\ \ \  \text{and} \ \  \
 {\cal C}_k \equiv \eta_{k}\hat Q^{\greater}_{\tS}
   -\eta_{\bar k}^{\ast}\hat Q^{\lesser}_{\tS}.
\end{split}
\ee
By setting $\dot\rho^{(n)}_{\bf n}=0$, we obtain
\begin{align}\label{DEOM_ss}
 0&=-{\cal L}^{(n)}_{\bf n}\bar\rho^{(n)}_{\bf n}
  +\sum_{k} n_k \Lambda_k \bar\rho^{(n)}_{{\bf n}^{+,-}_{k-1,k}}
\nl&\quad
  -i\sum_{k}{\cal A}\bar\rho^{(n+1)}_{{\bf n}_{k}^+}
  -i\sum_{k}\delta_{0p_k}n_{k}{\cal C}_k \bar\rho^{(n-1)}_{{\bf n}_{k}^-}.
\end{align}
Introduce an arbitrary parameter $\epsilon>0$ and recast the
above expression as
\begin{align}\label{SCI}
 \bar\rho^{(n)}_{\bf {n}}
&=\big({\cal L}^{(n)}_{\bf {n}}
  +\epsilon\big)^{-1}
 \Big[\epsilon\bar\rho^{(n)}_{\bf {n}}
 +\sum_{k} n_k\Lambda_k \bar\rho^{(n)}_{{\bf n}^{+,-}_{k-1,k}}
\nl&\quad\
 -i\sum_{k}{\cal A}\bar\rho^{(n+1)}_{\bf {n_{k}^+}}
 -i\sum_{k} \delta_{0p_k}n_{k}{\cal C}_k
   \bar\rho^{(n-1)}_{\bf {n_{k}^-}}
 \Big].
\end{align}
This constitutes a blocked Jacobi iteration scheme,
using the right-hand-side (rhs) DDOs to update the left-hand-side (lhs) one.
The SCI accommodates further the hierarchy structure
and the efficient on-the-fly filtering algorithm.\cite{Shi09084105}
It proceeds recursively with increasing $n$
in each iteration cycle, so that all
$\bar\rho_{{\bf n}_{k}^{-}}^{(n-1)}$
had just been updated.
This feature is also true for
some $\bar\rho^{(n)}_{{\bf n}^{+,-}_{k-1,k}}$,
the new ingredients of this work,
since the configurations in each individual
$n$--dissipatons subset are also pre-ordered.
Each hierarchy iteration cycle starts with the reduced system density matrix
$\rho^{\rm eq}_{\tS} \equiv \bar\rho^{(0)}_{\bf 0}$.
This determines also the common scaling parameter
for all DDOs in that cycle via the normalization constraint
on $\text{tr}_{\tS}\rho^{\rm eq}_{\tS}=1$.

 The SCI converges as long as the diagonal
part of $({\cal L}^{(n)}_{\bf n}+\epsilon)$
dominates.\cite{Zha17044105}
This suggests the lower bound for the SCI
relaxation parameter $\epsilon$.
Increasing $\epsilon$ will increase the numerical stability, but decrease the convergence speed.
For a good balance between accuracy and efficiency,
we choose $\epsilon\gtrsim\max(\Omega_{\tS},\sqrt{L\eta_{\max}})$, where $\Omega_{\tS}$ is the system spectrum span,
$\eta_{\max}=\max(|\eta_1|,\cdots,|\eta_K|)$
and $L$ the level of truncation.
The advantages of SCI scheme over
the Krylov subspace methods are remarkable.\cite{Zha17044105}

\section{The imaginary--time formalism}
\label{thappB}

 In line with $A_\text{hyb}$ as defined in \Eq{Ahyb_0},
the canonical partition function of the total
system--and--bath composite reads
$ Z_{\T} \equiv {\rm Tr}\, e^{-\beta H_{\T}}
=  Z_0 Z_\text{hyb}$.
Note that
$H_{\T}=H_{\tS}+h_{\B}+H_{\tS\B}$ with $[H_{\tS},h_{\B}]=0$.
The unhybrid part,  $Z_0\equiv {\rm Tr}\, e^{-\beta(H_{\tS}+h_{\B})}
= ({\rm tr}_{\tS} e^{-\beta H_{\tS}})({\rm tr}_{\B} e^{-\beta h_{\B}})
\equiv Z^{\tS}_0Z^{\B}_0$,
contributes to $A_0$ in \Eq{Ahyb_0} and
is treated as the reference.
The imaginary--time $i$-DEOM to be developed
in this appendix focuses on
\be\label{ZSB}
 {\rm tr}_{\tS}\varrho^{(0)}_{\bf 0}(\beta)
 ={\rm Tr}\hat\varrho_{\T}(\beta)=Z_\text{hyb} = e^{-\beta A_{\rm hyb}}.
\ee
Remarkably, the $i$-DDOs, which are
the $i$-DEOM dynamical variables,
acquire also the generic form of \Eq{DDOs}:
\be\label{iDDOs}
 \varrho^{(n)}_{\bf n}(\tau) ={\rm tr}_{\B}\big[
\big(\ti f_{K}^{n_K}\cdots\ti f_{1}^{n_1}\big)^{\circ}
\hat\varrho_{\T}(\tau)\big].
\ee
We will identity the $i$-dissipatons, $\{\ti f_{k}\}$,
the underlying dissipaton algebra and the resultant
$i$-DEOM formalism.
All these can be carried out
in parallel to \Sec{thsec2} and \Sec{thsec3},
but with the methods of analytical continuation.

 Let us start with the total composite $\hat\varrho_{\T}(\tau)$ in \Eq{iDDOs}.
In contact with the target \Eq{ZSB}, we have
\be\label{varrhoT_tau}
 \hat\varrho_{\T}(\tau)\equiv
 e^{-\tau H_{\T}}e^{-(\beta-\tau)(H_{\tS}+h_{\B})}
\big/(Z^{\tS}_0Z^{\B}_0).
\ee
It satisfies [\cf\Eq{Algrter} for notations]
\be\label{iLiou_eq}
 \frac{\d}{\d\tau}{\hat\varrho}_{\T}(\tau)
=-\big(H^{\times}_{\tS}+h^{\times}_{\B}
  +H^{\greater}_{\tS\B}\big)\hat\varrho_{\T}(\tau),
\ee
where $H_{\SB}=\hat Q_{\tS}\hat F_{\B}$.
The corresponding $i$-DDOs dynamic equation is then
[\cf\Eqs{DEOM}--(\ref{DDO_FB})]
\begin{align}\label{iDEOM}
 \frac{\d}{\d\tau}\varrho^{(n)}_{\bf n}(\tau)
&=-H^{\times}_{\tS}\varrho^{(n)}_{\bf n}(\tau)
-\varrho^{(n)}_{\bf n}(\tau;h^{\times}_{\B})
\nl&\quad
-\hat Q_{\tS}\varrho^{(n)}_{\bf n}(\tau;\hat F_{\B}^{\greater}),
\end{align}
with $H^{\times}_{\tS}\varrho^{(n)}_{\bf n}(\tau)
\equiv [H_{\tS},\varrho^{(n)}_{\bf n}(\tau)]$ and
\be\label{iDDO_hBF}
\begin{split}
 \varrho^{(n)}_{\bf n}(\tau;h^{\times}_{\B})
&\equiv {\rm tr}_{\B}\Big[
  \big(\ti f_{K}^{n_K}\cdots\ti f_{1}^{n_1}\big)^{\circ}
 \hat h^{\times}_{\B}\hat\varrho_{\T}(\tau)\big],
\\
 \varrho^{(n)}_{\bf n}(\tau;\hat F^{\greater}_{\B})
&\equiv{\rm tr}_{\B}\big[
  \big(\ti f_{K}^{n_K}\cdots\ti f_{1}^{n_1}\big)^{\circ}
 \hat F^{\greater}_{\B}\hat\varrho_{\T}(\tau)\big].
\end{split}
\ee
Moreover,  $\hat\varrho_{\T}(0)=e^{-\beta(H_{\tS}+h_{\B})}\big/(Z^{\tS}_0Z^{\B}_0)$
via \Eq{varrhoT_tau}. This determines
the initial $i$-DDOs the values of
\be\label{iDDOs0}
\varrho^{(0)}_{\bf 0}(0)=e^{-\beta H_{\tS}}/Z^{\tS}_0
\ \ \text{and} \ \
 \varrho_{\bf n}^{(n>0)}(0)=0.
\ee

 It is worth noting that
$h^{\times}_{\B}$ denotes the bare--bath $h_{\B}$--commutator
action. This validates the required analytical
continuation methods by setting $t=-i\tau$
to reach at the final $i$--DEOM formalism.
First of all, it acquires
\be\label{wtiF_tau_def}
 \wti{F}_{\B}(\tau) \equiv \hat F_{\B}(t=-i\tau)
 = e^{\tau h_{\B}}\hat F_{\B} e^{-\tau h_{\B}}.
\ee
We have $\la\wti{F}_{\B}(0)\wti{F}_{\B}(\tau)\ra_{\B}
=\la\wti{F}_{\B}(-\tau)\wti{F}_{\B}(0)\ra_{\B}$ and also
\be\label{iCorrelation}
 \la\wti{F}_{\B}(\tau)\wti{F}_{\B}(0)\ra_{\B}
=\la\wti{F}_{\B}(\beta - \tau)\wti{F}_{\B}(0)\ra_{\B}.
\ee
The corresponding \Eq{FDT} with $t=-i\tau$ reads\cite{Wei12}
\be\label{iFDT}
 \la\wti{F}_{\B}(\tau)\wti{F}_{\B}(0)\ra_{\B}
=\frac{1}{\pi}\!\int_{-\infty}^{\infty}\!{\rm d}\w\,
\frac{e^{-\tau\w}J(\w)}{1-e^{-\beta\w}}.
\ee
This is an analytical function in
the required $\tau\in [0,\beta]$.
Consequently, the Cauchy's contour integration
technique is applicable.
On the other hand, one can recast \Eq{iFDT} as
\[
 \la\wti{F}_{\B}(\tau)\wti{F}_{\B}(0)\ra_{\B}
=\lim_{t\rightarrow 0}\frac{1}{\pi}\!\int_{-\infty}^{\infty}\!{\rm d}\w\,
\frac{e^{-i\w(t-i\tau)}J(\w)}{1-e^{-\beta\w}}.
\]
The analytical continuation of \Eq{FF_FSD} reads then
\bsube\label{iFF_FSD}
\begin{align}
 \la\wti{F}_{\B}(\tau)\wti{F}_{\B}(0)\ra_{\B}
&=\sum_{k=1}^K\eta_{k}\phi_{k}(-i\tau),
\label{iFF_FSD1}\\
 \la\wti{F}_{\B}(0)\wti{F}_{\B}(\tau)\ra_{\B}
&=\sum_{k=1}^K \eta^{\ast}_{\bar k}\phi_{k}(-i\tau),
\label{iFF_FSD2}
\end{align}
\esube
with $\phi_{k}(-i\tau)=(-i\gamma_k\tau)^{p_k}e^{i\gamma_{k}\tau}$,
\Eq{phi_FSD}.
We have numerically verified
that in the limit of $K\rightarrow\infty$,
\Eq{iFF_FSD1} agrees perfectly with
\Eq{iFDT} where $\tau\in [0,\beta]$.
The resultant sum is practically real,
which together with $\la\wti{F}_{\B}(0)\wti{F}_{\B}(\tau)\ra_{\B}
=\la\wti{F}_{\B}(-\tau)\wti{F}_{\B}(0)\ra_{\B}$
give rise to \Eq{iFF_FSD2}.
The associate index $\bar k$ was defined before after \Eq{FF_in_phi}.
To reproduce \Eq{iFF_FSD}, we identify [\cf\Eqs{F_f} and (\ref{ff_t})]
\begin{align}\label{iF_f}
 \hat F_{\B} &= \sum_{k=1}^K \ti{f}_{k},
\intertext{with}
\label{iff_t}
\begin{split}
 \la\ti f_{k}(\tau)\ti f_j(0)\ra_{\B}
&=\delta_{kj}\eta_k\phi_k(-i\tau),
\\
 \la\ti f_j(0)\ti f_{k}(\tau)\ra_{\B}
&=\delta_{kj}\eta^{\ast}_{\bar k}\phi_k(-i\tau).
\end{split}
\end{align}
As defined in \Eq{wtiF_tau_def},
 $\ti{f}_{k}(\tau) = e^{\tau h_{\B}}\ti{f}_{k} e^{-\tau h_{\B}}$
that satisfies
$-(\partial \ti f_k/\partial\tau)_{\B}=[\hat f_{\B}, \hat h_{\B}]$.
We obtain [\cf\Eq{diff}]
\begin{align*}
 \varrho_{\bf n}^{(n)}(\tau;h^{\times}_{\B})
=i\sum_{k} n_{k}\gamma_{k}
   \bigg[\frac{p_{k}\eta_{k}}{\eta_{k-1}}
    \varrho^{(n)}_{{\bf n}^{+,-}_{k-1,k}}\!(\tau)
  -\varrho^{(n)}_{\bf n}(\tau)\bigg],
\intertext{and also [\cf\Eq{Wick}]}
 \varrho^{(n)}_{\bf n}(\tau;\hat F^{\greater}_{\B})
=\sum_{k}
   \left[\varrho^{(n+1)}_{{\bf n}_{k}^{+}}(\tau)
  +\delta_{0 p_{k}} n_{k}\eta_{k}
   \varrho^{(n-1)}_{{\bf n}^{-}_{k}}(\tau)\right].
\end{align*}
Both are identical to their real--time counterparts.
 All parameters are identical to
to those specified in \Sec{thsec2B}.
The resultant \Eq{iDEOM} reads [\cf\Eqs{calLn_def} and (\ref{DEOM_ss})]
\begin{align}\label{iDEOM_final}
 \frac{\d\varrho^{(n)}_{\bf n}}{\d\tau}
&=i{\cal L}^{(n)}_{\bf n}\varrho^{(n)}_{\bf n}
 -i\sum_{k} n_{k}\Lambda_k
    \varrho^{(n)}_{{\bf n}^{+,-}_{k-1,k}}
\nl&\quad
-\hat Q_{\tS}\sum_{k}
   \left(\varrho^{(n+1)}_{{\bf n}_{k}^{+}}
  +\delta_{0 p_{k}} n_{k}\eta_{k}
   \varrho^{(n-1)}_{{\bf n}^{-}_{k}}\right).
\end{align}
This is the unified $i$-DEOM formalism of this work,
with the initial values of \Eq{iDDOs0}
to be propagated to $\{\varrho^{(n)}_{\bf n}(\tau=\beta); n=0,\cdots, L\}$.
Together with \Eq{ZSB}, we can evaluate
$A_{\rm hyb}(T)$
without invoking the $\lambda$--integration as \Eq{Ahyb_integral}.
 Moreover, the above $i$-dissipaton formalism
goes with a one--to--one correspondence manner,
such as \Eq{iDDOs} versus \Eq{DDOs}  and \Eq{ff_t} versus \Eq{iff_t},
respectively. As results,
\be\label{barrhon_varrhon}
 \bar\rho^{(n)}_{\bf n}(T)
=\frac{\varrho^{(n)}_{\bf n}(\beta)}{Z_\text{hyb}}
=\frac{\varrho^{(n)}_{\bf n}(\beta)}{{\rm tr}_{\tS}\varrho^{(0)}_{\bf 0}(\beta)}.
\ee
This is an alternative approach to
the equilibrium DDOs, \Eq{rhon_eq} or \Eq{DEOM_ss}.


\end{appendix}


\begin{thebibliography}{10}

\bibitem{Hu922843}
B.~L. Hu, J.~P. Paz, and Y.~Zhang,
\newblock Phys. Rev. D {\bf 45}, 2843 (1992).

\bibitem{Xu09074107}
R.~X. Xu, B.~L. Tian, J.~Xu, and Y.~J. Yan,
\newblock J. Chem. Phys. {\bf 130}, 074107 (2009).

\bibitem{Cal83374}
A.~O. Caldeira and A.~J. Leggett,
\newblock Ann. Phys. {\bf 149}, 374 (1983),
\newblock [Erratum: {\bf 153}, 445 (1984)].

\bibitem{Gra8487}
H.~Grabert, U.~Weiss, and P.~Talkner,
\newblock Z. Phys. B Condens. Matter {\bf 55}, 87 (1984).

\bibitem{Gra88115}
H.~Grabert, P.~Schramm, and G.~L. Ingold,
\newblock Phys. Rep. {\bf 168}, 115 (1988).

\bibitem{Hor057325}
C.~H{\"{o}}rhammer and H.~B{\"{u}}ttner,
\newblock J. Phys. A. Math. Gen. {\bf 38}, 7325 (2005).

\bibitem{Hor081161}
C.~H{\"{o}}rhammer and H.~B{\"{u}}ttner,
\newblock J. Stat. Phys. {\bf 133}, 1161 (2008).

\bibitem{OCo0615}
R.~F. O'Connell,
\newblock J. Stat. Phys. {\bf 124}, 15 (2006).

\bibitem{Cam09392002}
M.~Campisi, P.~Talkner, and P.~H{\"{a}}nggi,
\newblock J. Phys. A: Math. Theor. {\bf 42}, 392002 (2009).

\bibitem{Fey63118}
R.~P. Feynman and F.~L. \mbox{Vernon, Jr.},
\newblock Ann. Phys. {\bf 24}, 118 (1963).

\bibitem{Tan89101}
Y.~Tanimura and R.~Kubo,
\newblock J. Phys. Soc. Jpn. {\bf 58}, 101 (1989).

\bibitem{Tan906676}
Y.~Tanimura,
\newblock Phys. Rev. A {\bf 41}, 6676 (1990).

\bibitem{Tan06082001}
Y.~Tanimura,
\newblock J. Phys. Soc. Jpn. {\bf 75}, 082001 (2006).

\bibitem{Yan04216}
Y.~A. Yan, F.~Yang, Y.~Liu, and J.~S. Shao,
\newblock Chem. Phys. Lett. {\bf 395}, 216 (2004).

\bibitem{Xu05041103}
R.~X. Xu, P.~Cui, X.~Q. Li, Y.~Mo, and Y.~J. Yan,
\newblock J. Chem. Phys. {\bf 122}, 041103 (2005).

\bibitem{Xu07031107}
R.~X. Xu and Y.~J. Yan,
\newblock Phys. Rev. E {\bf 75}, 031107 (2007).

\bibitem{Tan14044114}
Y.~Tanimura,
\newblock J. Chem. Phys. {\bf 141}, 044114 (2014).

\bibitem{Tan15144110}
Y.~Tanimura,
\newblock J. Chem. Phys. {\bf 142}, 144110 (2015).

\bibitem{Yan14054105}
Y.~J. Yan,
\newblock J. Chem. Phys. {\bf 140}, 054105 (2014).

\bibitem{Yan16110306}
Y.~J. Yan, J.~S. Jin, R.~X. Xu, and X.~Zheng,
\newblock Frontiers Phys. {\bf 11}, 110306 (2016).

\bibitem{Zha18780}
H.~D. Zhang, R.~X. Xu, X.~Zheng, and Y.~J. Yan,
\newblock Mol. Phys. {\bf 116}, 780 (2018),
\newblock Special Issue, ``Molecular Physics in China''.

\bibitem{Wan20041102}
Y.~Wang, R.~X. Xu, and Y.~J. Yan,
\newblock J. Chem. Phys. {\bf 152}, 041102 (2020).

\bibitem{Zha15024112}
H.~D. Zhang, R.~X. Xu, X.~Zheng, and Y.~J. Yan,
\newblock J. Chem. Phys. {\bf 142}, 024112 (2015).

\bibitem{Zha16204109}
H.~D. Zhang, Q.~Qiao, R.~X. Xu, and Y.~J. Yan,
\newblock J. Chem. Phys. {\bf 145}, 204109 (2016).

\bibitem{Jin15234108}
J.~S. Jin, S.~K. Wang, X.~Zheng, and Y.~J. Yan,
\newblock J. Chem. Phys. {\bf 142}, 234108 (2015).

\bibitem{Jin18043043}
J.~S. Jin, S.~K. Wang, J.~H. Zhou, W.~M. Zhang, and Y.~J. Yan,
\newblock New. J. Phys. {\bf 20}, 043043 (2018).

\bibitem{Jin20235144}
J.~S. Jin,
\newblock Phys. Rev. B {\bf 101}, 235144 (2020).

\bibitem{Cui19024110}
L.~Cui, H.~D. Zhang, X.~Zheng, R.~X. Xu, and Y.~J. Yan,
\newblock J. Chem. Phys. {\bf 151}, 024110 (2019).

\bibitem{Zha20064107}
H.~D. Zhang, L.~Cui, H.~Gong, R.~X. Xu, X.~Zheng, and Y.~J. Yan,
\newblock J. Chem. Phys. {\bf 152}, 064107 (2020).

\bibitem{Kir35300}
J.~G. Kirkwood,
\newblock J. Chem. Phys. {\bf 3}, 300 (1935).

\bibitem{Shu71413}
R.~C. Shuela and E.~R. Muller,
\newblock Phys. Stat. Sol. (b) {\bf 43}, 413 (1971).

\bibitem{Zon08041103}
R.~van Zon, L.~Hern\'andez de~la Pe\~na, G.~H. Peslherbe, and J.~Schofield,
\newblock Phys. Rev. E {\bf 78}, 041103 (2008).

\bibitem{Zon08041104}
R.~van Zon, L.~Hern\'andez de~la Pe\~na, G.~H. Peslherbe, and J.~Schofield,
\newblock Phys. Rev. E {\bf 78}, 041104 (2008).

\bibitem{Wei12}
U.~Weiss,
\newblock {\em Quantum Dissipative Systems},
\newblock World Scientific, Singapore, 2012,
\newblock 4$^{\rm rd}$ ed.

\bibitem{Yan05187}
Y.~J. Yan and R.~X. Xu,
\newblock Annu. Rev. Phys. Chem. {\bf 56}, 187 (2005).

\bibitem{Zhe121129}
X.~Zheng, R.~X. Xu, J.~Xu, J.~S. Jin, J.~Hu, and Y.~J. Yan,
\newblock Prog. Chem. {\bf 24}, 1129 (2012).

\bibitem{Xu18114103}
R.~X. Xu, Y.~Liu, H.~D. Zhang, and Y.~J. Yan,
\newblock J. Chem. Phys. {\bf 148}, 114103 (2018).

\bibitem{Liu18245}
Y.~Liu, R.~X. Xu, H.~D. Zhang, and Y.~J. Yan,
\newblock Chin. J. Chem. Phys. {\bf 31}, 245 (2018).

\bibitem{Hu10101106}
J.~Hu, R.~X. Xu, and Y.~J. Yan,
\newblock J. Chem. Phys. {\bf 133}, 101106 (2010).

\bibitem{Hu11244106}
J.~Hu, M.~Luo, F.~Jiang, R.~X. Xu, and Y.~J. Yan,
\newblock J. Chem. Phys. {\bf 134}, 244106 (2011).

\bibitem{Din12224103}
J.~J. Ding, R.~X. Xu, and Y.~J. Yan,
\newblock J. Chem. Phys. {\bf 136}, 224103 (2012).

\bibitem{Din16204110}
J.~J. Ding, H.~D. Zhang, Y.~Wang, R.~X. Xu, X.~Zheng, and Y.~J. Yan,
\newblock J. Chem. Phys. {\bf 145}, 204110 (2016).

\bibitem{Tan15224112}
Z.~F. Tang, X.~L. Ouyang, Z.~H. Gong, H.~B. Wang, and J.~L. Wu,
\newblock J. Chem. Phys. {\bf 143}, 224112 (2015).

\bibitem{Dua17214308}
C.~R. Duan, Z.~F. Tang, J.~S. Cao, and J.~L. Wu,
\newblock Phys. Rev. B {\bf 95}, 214308 (2017).

\bibitem{Din17024104}
J.~J. Ding, Y.~Wang, H.~D. Zhang, R.~X. Xu, X.~Zheng, and Y.~J. Yan,
\newblock J. Chem. Phys. {\bf 146}, 024104 (2017).

\bibitem{Ye17074111}
L.~Z. Ye, H.~D. Zhang, Y.~Wang, X.~Zheng, and Y.~J. Yan,
\newblock J. Chem. Phys. {\bf 147}, 074111 (2017).

\bibitem{Zha17044105}
H.~D. Zhang, Q.~Qiao, R.~X. Xu, X.~Zheng, and Y.~J. Yan,
\newblock J. Chem. Phys. {\bf 147}, 044105 (2017).

\bibitem{Phi051781}
J.~C. Phillips, R.~Braun, W.~Wang, J.~Gumbart, E.~Tajkhorshid, E.~Villa,
  C.~Chipot, R.~D. Skeel, L.~Kal\'{e}, and K.~Schulten,
\newblock J. Comput. Chem. {\bf 26}, 1781 (2005).

\bibitem{Sal13198}
R.~Salomon-Ferrer, D.~A. Case, and R.~C. Walker,
\newblock Wiley Interdisciplinary Reviews: Comput. Mol. Sci. {\bf 3}, 198
  (2013).

\bibitem{Abr1519}
M.~J. Abraham, T.~Murtola, R.~Schulz, S.~Pall, J.~C. Smith, B.~Hess, and
  E.~Lindahl,
\newblock SoftwareX {\bf 1-2}, 19  (2015).

\bibitem{Hir862521}
J.~E. Hirsch and R.~M. Fye,
\newblock Phys. Rev. Lett. {\bf 56}, 2521 (1986).

\bibitem{Gul11349}
E.~Gull, A.~J. Millis, A.~I. Lichtenstein, A.~N. Rubtsov, M.~Troyer, and
  P.~Werner,
\newblock Rev. Mod. Phys. {\bf 83}, 349 (2011).

\bibitem{Whi922863}
S.~R. White,
\newblock Phys. Rev. Lett. {\bf 69}, 2863 (1992).

\bibitem{Vid03147902}
G.~Vidal,
\newblock Phys. Rev. Lett. {\bf 91}, 147902 (2003).

\bibitem{Wil75773}
K.~G. Wilson,
\newblock Rev. Mod. Phys. {\bf 47}, 773 (1975).

\bibitem{Bul08395}
R.~Bulla, T.~A. Costi, and T.~Pruschke,
\newblock Rev. Mod. Phys. {\bf 80}, 395 (2008).

\bibitem{Saa03}
Y.~Saad,
\newblock {\em Iterative Methods for Sparse Linear Systems},
\newblock Society for Industrial and Applied Mathematics, xx, 2nd edition,
  2003.

\bibitem{Shi09084105}
Q.~Shi, L.~P. Chen, G.~J. Nan, R.~X. Xu, and Y.~J. Yan,
\newblock J. Chem. Phys. {\bf 130}, 084105 (2009).

\end{thebibliography}

\end{document}